\documentclass[twocolumn,superscriptaddress,aps,pra]{revtex4}
\usepackage{amsmath,amsthm,amssymb}
\usepackage{latexsym}
\usepackage{amscd,graphicx}
\usepackage[titletoc]{appendix}
\usepackage{epsfig}
\usepackage{epstopdf}
\usepackage[normalem]{ulem}
\usepackage[dvipsnames]{xcolor}
\usepackage[colorlinks=true,linkcolor=red,citecolor=blue]{hyperref}
\begin{document}
	
	\newcommand{\cmt}[1]{{\textcolor{red}{[#1]}}}
	\newcommand{\qn}[1]{{\textcolor{red}{ (?)  #1 }}}
	\newcommand{\chk}[1]{{\textcolor{green}{#1}}}
	\newcommand{\del}[1]{{\textcolor{blue}{ \sout{#1}}}}
	\newcommand{\rvs}[1]{{\textcolor{blue}{#1}}}

\title{Strong single-photon optomechanical coupling in a hybrid quantum system}

\author{Jiaojiao Chen}
\affiliation{Department of Physics, Wenzhou University, Zhejiang 325035, China}
\affiliation{Hefei Preschool Education College, Hefei 230013, China}


\author{Zhuanxia Li}
\affiliation{Department of Physics, Wenzhou University, Zhejiang 325035, China}

\author{Xiao-Qing Luo}
\affiliation{Hunan Province KeyLaboratory for Ultra-Fast Micro/Nano Technology and Advanced Laser Manufacture, School of Electrical Engineering, University of SouthChina, Hengyang, 421001, China}

\author{Wei Xiong}

\altaffiliation{xiongweiphys@wzu.edu.cn}
\affiliation{Department of Physics, Wenzhou University, Zhejiang 325035, China}

\author{Mingfeng Wang}
\altaffiliation{mfwang@wzu.edu.cn}
\affiliation{Department of Physics, Wenzhou University, Zhejiang 325035, China}

\author{Hai-Chao Li}
\altaffiliation{hcl2007@foxmail.com}
\affiliation{College of Physics and Electronic Science, Hubei Normal University, Huangshi 435002, China}


\date{\today }

\begin{abstract}
Engineering strong single-photon optomechanical couplings is crucial for optomechanical systems. Here, we propose a hybrid quantum system consisting of a nanobeam (phonons) coupled to a spin ensemble and a cavity (photons) to overcome it. Utilizing the critical property of the lower-branch polariton (LBP) formed by the ensemble-phonon interaction, the LBP-cavity coupling can be greatly enhanced by three orders magnitude of the original one, while the  upper-branch polariton (UBP)-cavity coupling is fully suppressed. Our proposal breaks through the condition of the coupling strength less than the critical value in previous schemes using two harmonic oscillators.  Also, strong Kerr effect can be induced in our proposal. This shows our proposed approach can be used to study quantum nonlinear and nonclassical effects in weakly coupled optomechanical systems.
\end{abstract}


\maketitle

\section{Introduction}
Due to the potential applications in studying optomechanically induced transparency~\cite{Weis-2010,Teufel-2011,Xiong-2016}, optical bistability~\cite{Xiong-2016,Sarala-2015}, higher-orders sidebands~\cite{Jiao-2016,Suzuki-2015}, precision measurements~\cite{Rugar-2004,Li-2013,Xiong-2017}, gravitational wave detectors~\cite{Braginsky-2002,Abramovici-1992}, phonon lasers~\cite{Jing-2014,Grudinin-2010}, cavity optomechanics~\cite{Aspelmeyer-2014} has received great interest in the past decades. However, realizing strong single-photon optomechanical coupling is still challenging for most of optomechanical setups ~\cite{Groblacher-2009,Rocheleau-2010,Brennecke-2008,Eichenfield-2009} to date. 

One experimentally used method is to apply a strong driving field on an optomechanical cavity~\cite{Vitali-2007}, resulting in a strong linearized optomechanical coupling larger than the decay rate of the cavity. However, this method sacrifices its inherent nonlinearity. For this, various approaches are proposed for improving the single-photon optomechanical coupling, including mechanical oscillators arrays~\cite{Xuereb-2012}, squeezing effects~\cite{Lu-2015, Li-2016}, Kerr nonlinearity via the Josephson effect~\cite{Heikkila-2014,Liao-2014}, as well as a single atom~\cite{Neumeier-2018} and superconducting circuits~\cite{Via-2015,Shevchuk-2017,Kounalakis-2020,Haque-2020,Bothner-2021}. These remarkable schemes indicate that enhancing optomechanical coupling is possible using hybrid systems. In addition, quantum criticality is utilized to efficiently improve system coupling~\cite{Lu-2013,Xiong-2020}. In their proposals~\cite{Lu-2013,Xiong-2020}, the critical point (CP) can only be approached by unidirectionally increasing the photon-phonon coupling strength. However, it is out of work when the photon-phonon coupling exceeds the critical value. 

Motivated by this, we replace one of the optomechanical cavities in~\cite{Lu-2013} by a nitrogen vacancy ensemble (NVE) embedded in a diamond nanobeam. As well known, the NVE can be mapped to the bosonic mode in the large number limit~\cite{Hummer-2012, Qiu-2014, Zhang-2021}, so that the NVE can be regarded as a single-mode cavity. We consider here the case that the NVE-phonon coupling exceeds the critical value, so that the low-excitation for the ensemble is broken and higher order terms are included. This gives rise to a nonlinear NVE-phonon coupling in the thermodynamics limit, which can hybridize the NVE and the phonon as two polaritons. Utilizing the critical property of the lower-branch polariton (LBP) formed by the coupled NVE-phonon, the LBP-cavity coupling can be greatly enhanced by three orders magnitude of the original one, while the upper-branch polariton (UBP)-cavity coupling is fully suppressed. In such an effective polariton-cavity optomechanical system, strong Kerr effect can be introduced. This shows our proposed approach can be used to study quantum nonlinear and nonclassical effects in weakly coupled optomechanical systems~\cite{Lu-2013}.

The paper is organized as follows. In Sec. II, the model and the Hamiltonian are introduced. Then we study the linearized mangnon-phonon interaction Hamiltonian in Sec. III. In Sec. IV, enhancing optomechanical coupling is investigated. Finally, a conclusion is given in Sec. V.

\section{Model and Hamiltonian}

\begin{figure}
	\center
	\includegraphics[scale=.37]{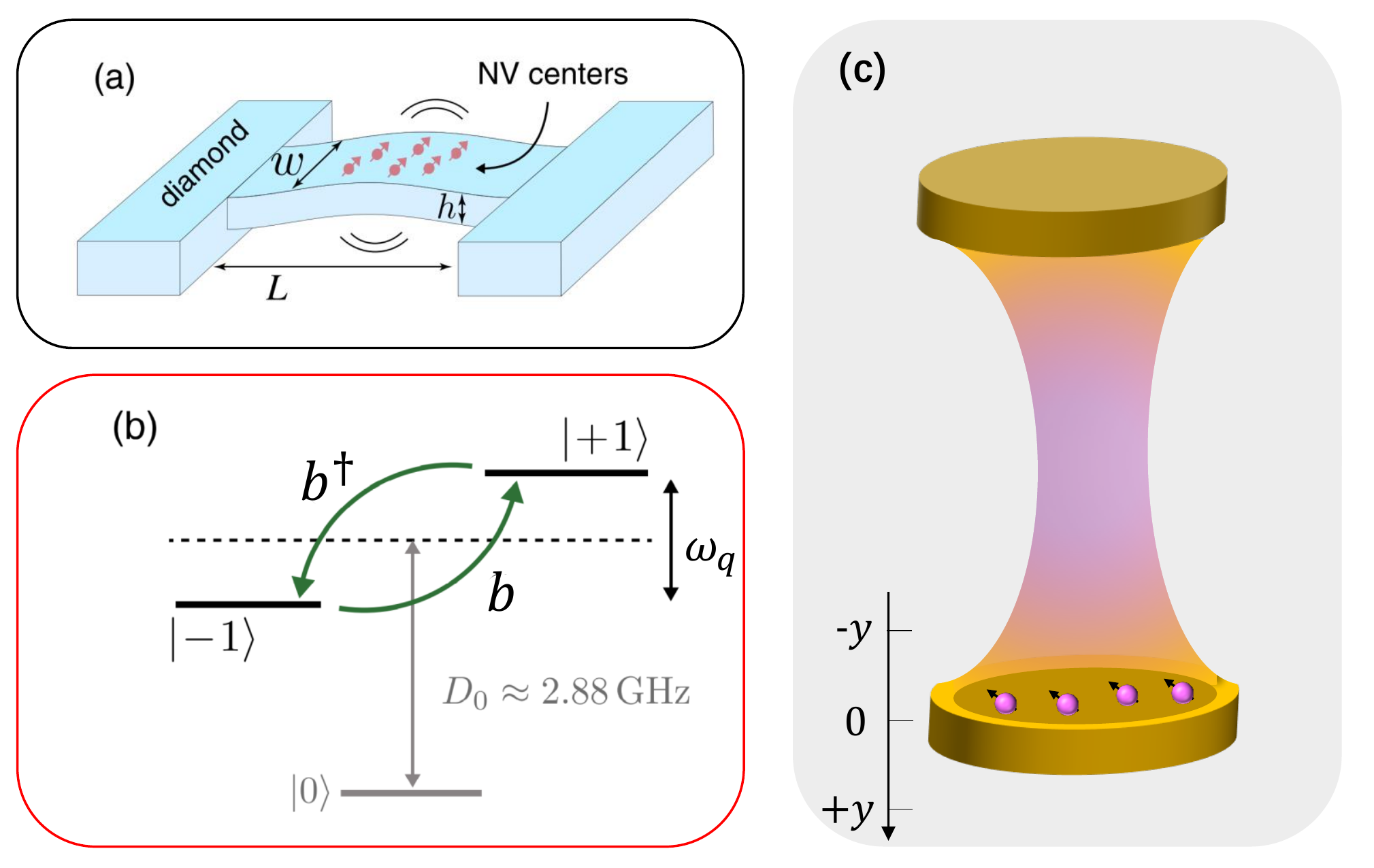}
	\caption{(a) The nitrogen vacancy centers are embedded in a diamond nanobeam~\cite{Bennett-2013}. (b) Spin triplet states of the NV electronic ground state. Local perpendicular strain induced by beam bending mixes the $|\pm1\rangle$ states. (c) The proposed hybrid system consisting of a nanobeam coupled to a spin ensemble and a cavity.}\label{fig1}
\end{figure}
We consider an ensemble consisting of $N$ NV center spins embedded in a crystal diamond nanobeam with frequency $\omega_m$, as shown in Fig.~\ref{fig1}(a). The electronic ground state of the NV center spin is $S=1$ triplet states labeled by $|m_s=0,\pm1\rangle$, as shown in Fig.~\ref{fig1}(b). In the presence of external electric field ${\bf E}$ and magnetic field ${\bf B}$, the Hamiltonian for a single NV spin is~\cite{Doherty-2012,Bennett-2013} (setting $\hbar=1$)
\begin{align}
H_{\rm NV}=&(D_0+d_\parallel E_z)S_z^2+\mu_B g_s {\bf S}\cdot {\bf B},\nonumber\\
&-d_\perp[E_x(S_xS_y+S_yS_x)+E_y(S_x^2-S_y^2)],\label{q1}
\end{align}
where $D_0/2\pi\approx2.88$~GHz is the zero field splitting, $g_s\approx2$, $\mu_B$ is the Bohr magneton, and $d_\parallel~(d_\perp)$ is the ground
state electric dipole moment in the direction parallel (perpendicular) to the NV axis~\cite{Vanoort-1990}. When the beam flexes, it strains the diamond lattice, which in turn couples directly to the spin triplet states in the NV electronic ground state. We here consider the case that the nanobeam is coupled to the transition between $|\pm1\rangle$ of the spin, with Zeeman splitting $\omega_q=\mu_B g_s B_z/\hbar$~[see Fig.~\ref{fig1}(b)]. The perpendicular component of strain $E_\perp$ hybridize the $|\pm1\rangle$ states. For small beam displacements, $E_\perp=E_0(b+b^\dag)$ is the perpendicular strain resulting from the zero point
motion of the nanobeam, where $b$ ($b^\dag$) is the annihilation (creation) operator of the beam.  The parallel component of strain shifts both states $|\pm1\rangle$  relative to $|0\rangle$~\cite{Acosta-2010}. This effect however can be efficiently suppressed in the near-resonant coupling, i.e., $\omega_q-\omega_m\ll D_0$. Thus,  the system Hamiltonian~(\ref{q1}) in the basis $|\pm\rangle$ can be written as
\begin{align}
H_{\rm NM}=&\omega_m b^\dag b+\omega_qJ_z+g(b+b^\dag)J_x,\label{q2}
\end{align}
where
\begin{align}
J_+=\sum\limits_{i=1}^{N}\sigma_+^{(i)},~J_-=\sum\limits_{i=1}^{N}\sigma_-^{(i)},~J_z=\frac{1}{2}\sum\limits_{i=1}^{N}\sigma_z^{(i)}, \label{q3}
\end{align}
are collective spin operators with commutation relations $[J_+,J_-]=2J_z$ and $[J_z,J_\pm]=\pm J_z$. Here  $\sigma_+^{(i)}=[\sigma_-^{(i)}]^\dag=|1\rangle_i\langle -1|+|-1\rangle_i\langle 1|$ and $\sigma_z^{(i)}=|1\rangle_i\langle 1|-|-1\rangle_i\langle -1|$ are Pauli operators for the $i$th two-level spin. For simplicity, these spins are assumed to be identical, such that they have the same transition frequency $\omega_q$ and their transitions from the ground state $|-1\rangle$ to
the excited state $|+1\rangle$ are driven by the same coupling $g$ to the nanobeam. 

We now consider the case of $N\rightarrow\infty$, the dynamics of the collective spin can be mapped to a bosonic mode $c$ via the Holstein-Primakoff transformation~\cite{Holstein-1940}, i.e.,
\begin{align}
J_-=&\sqrt{N}\xi c,~~
J_+=\sqrt{N}c^\dag\xi,~~
J_z=c^\dag c-N/2.\label{q4}
\end{align}
Then, Eq.~(\ref{q2}) reduces to
\begin{align}
H_{\rm NMP}=\omega_m b^\dag b+\omega_q c^\dag c+G(b+b^\dag)(\xi c+c^\dag \xi),\label{q5}
\end{align}
where $G=g\sqrt{N}$ is the effective enhanced coupling between the NVE and the nanobeam, and $\xi=\sqrt{1-c^\dag c/2N}$ is nonlinear dependence of the excitation number $c^\dag c$, characterizing the transition from the linear to nonlinear coupling between the nanobeam and the NVE (excited spins). For example, $\xi=1$ means the coupling between the NVE and the nanobeam is bilinear, while it is nonlinear for $\xi\neq1$. 

\section{The effective linearized Hamiltonian for NVE coupled to the nanobeam}
When the effective coupling strength between the NVE and the nanobeam in Eq.~(\ref{q5}) is strong, generally exceeding the critical coupling strength, i.e., $G\geq G_{c}=\sqrt{\omega_m\omega_q}/2$, each operator of the system can acquire a macroscopic displacement. This is due to the fact that the strong  coupling results in more spins excited in the thermodynamics limit $N\rightarrow\infty$, especially for the large number of the NV spins~($\langle c^\dag c\rangle\gg1$).  Therefore, we write each operator in Eq.~(\ref{q5}) into the expectation value plus the fluctuation, i.e., 
\begin{align}\label{q6}
b\rightarrow&b+\sqrt{\alpha_b},~~
c\rightarrow c-\sqrt{\alpha_c},
\end{align}
and we have
\begin{align}\label{q7}
H_{\rm NMP}=&\omega_m b^\dag b+\omega_q c^\dag c+\omega_m\alpha_b+\omega_q(\alpha_c-N/2)\notag\\
&+\omega_m\sqrt{\alpha_b}(b^\dag+b)-\omega_q\sqrt{\alpha_c}(c^\dag+c)\notag\\
&+G\sqrt{\frac{k}{N}}(b^\dag+b+2\sqrt{\alpha_b})(c^\dag\xi_1+\xi_1 c-2\sqrt{\alpha_c}\xi_1).
\end{align}
where $k=N-\alpha_c$ and the parameter $\xi$ is corrected as
\begin{align}
\xi_1=\sqrt{1-\frac{c^\dag c-\sqrt{\alpha_c}(c^\dag+c)}{k}}.\label{q8}
\end{align}
In the thermodynamic limit ($\langle c^\dag c\rangle/N\ll 1$), $\xi_1$ can be expanded to its second order in $c^\dag c/N$. Substituting this expansion into Eq.~(\ref{q7}) and keeping the terms involving single- and two-body operators, the Hamiltonian in Eq.~(\ref{q7}) can be approximately written as
\begin{align}
H_{\rm NMP}\approx&\omega_m b^\dag b+\Omega_q c^\dag c+\mathcal{E}_b(b^\dag+b)+\mathcal{E}_c(c^\dag+c)\nonumber\\
&+\mathcal{G}(b^\dag+b)(c^\dag+c)+\eta(c^\dag+c)^2.\label{q9}
\end{align}
Here, $\Omega_q=\omega_q+2G\sqrt{\alpha_b\alpha_c/Nk}$ is the effective frequency of the mangnon. $\mathcal{E}_b=\omega_m\sqrt{\alpha_b}-2G\sqrt{k\alpha_c/N}$ and $\mathcal{E}_c=-\omega_q\sqrt{\alpha_c}+4G\sqrt{k\alpha_b/N}(N/2-\alpha_c)$ are amplitudes of the effective driving fields. $\mathcal{G}=2G(N/2-\alpha_c)/\sqrt{Nk}$ is the effective coupling strength between the NVE and the nanobeam. $\eta=(G/2k)\sqrt{\alpha_b\alpha_c/Nk}(2k+\alpha_c)$ is the effective coefficient for generating a pair of degenerate NVEs. The effective driving field can be eliminated when
\begin{align}
\mathcal{E}_b=\mathcal{E}_c=0.\label{q10}
\end{align}
This condition directly gives rise to
\begin{align}
\sqrt{\alpha_b}=\frac{2G}{\omega_m}\sqrt{\frac{N}{4}(1-\mu^2)},~~ \sqrt{\alpha_c}=\sqrt{\frac{N}{2}(1-\mu)},\label{q11}
\end{align}
where
\begin{align}
\mu={G_c^2}/{G^2}\leq1 \label{q12}
\end{align}
is the redefined critical paramter. Keeping the condition in Eq.~(\ref{q10}) satisfied, Eq.~(\ref{q9}) can be reduced to
\begin{align}
H_{\rm eff}=&\omega_m b^\dag b+\Omega_q c^\dag c+\mathcal{G}(b^\dag+b)(c^\dag+c)+\eta(c^\dag+c)^2,\label{q13}
\end{align}
which is the linearized Hamiltonian between the NVE and the nanobeam. Using the new critical parameter $\mu$, the parameters in Eq.~(\ref{q9}) or q.~(\ref{q13}) can be rewritten as
\begin{align}
\Omega_q=&\omega_q\frac{1+\mu}{2\mu},\nonumber\\ \mathcal{G}=&G\mu\sqrt{\frac{2}{1+\mu}},\nonumber\\ \eta=&\omega_q\frac{(1-\mu)(3+\mu)}{8\mu(1+\mu)}.
\end{align}

\section{Quantum criticality}

For the condition in Eq.~(\ref{q12}), the equality ($\mu=1$) corresponds to $\alpha_b=\alpha_c=0$, resulting in $\Omega_q=\omega_q$, $\mathcal{G}=G$ and $\eta=0$, so Eq.~(\ref{q13}) becomes
\begin{align}
H_{\rm eff}^{(0)}=\omega_m b^\dag b+\omega_q c^\dag c+G(b^\dag+b)(c^\dag+ c),\label{q14}
\end{align}
which describes a bilinear interaction between the NVE and the nanobeam, stemming from the zero order approximation of $\xi_1$ in Eq.~(\ref{q8}) or the low-excitation approximation. Such a typical Hamiltonian can have a critical property \cite{Emary1-2003,Emary2-2003} by increasing the coupling strength $G$ to the critical coupling strength $G_c$. Specifically, the eigen frequency of the LBP $\omega_{\rm LB}$ can transit from the complex to real numbers with increasing the coupling strength $G$. At the CP $G=G_c=\sqrt{\omega_m\omega_b}/2$,~$\omega_{\rm LB}=0$. Such critical properties can be used to enhance the single-photon optomechanical coupling~\cite{Lu-2013} and the polariton mediated spin-spin coupling~\cite{Xiong-2020}. However, the enhancement in their proposals~\cite{Lu-2013,Xiong-2020}  can only be realized in the region of $G<G_c$.

\begin{figure}
	\center
	\includegraphics[scale=0.45]{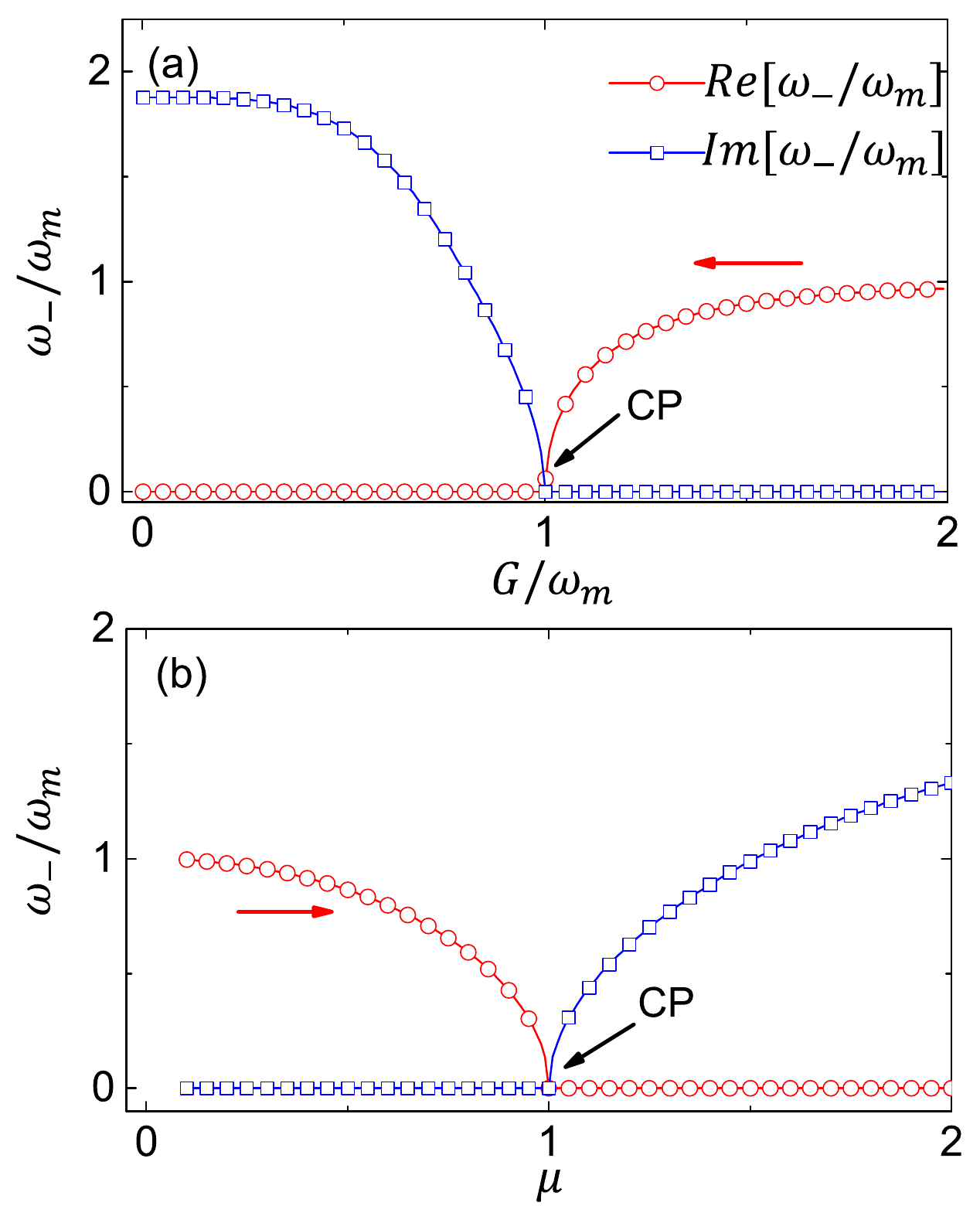}
	\caption{The normalized eigen frequency of the LBP, given by Eq.~(\ref{q13}) or (\ref{q16}), as functions of the normalized coupling strength $G/\omega_m$ and $\mu$ for $\omega_b/\omega_m=1$ and $10$. Here $\omega_b/\omega_m$=4 is chosen.}\label{fig2}
\end{figure}

For the case of $\mu<1$ in Eq.~(\ref{q12}), the system Hamiltonian is governed by Eq.~(\ref{q13}), which can be fully diagonalized via a Bogoliubov transformation~\cite{Emary1-2003}, as
\begin{align}
H_{\rm Diag}=\omega_+ d_+^\dag d_++\omega_- d_-^\dag d_-,\label{q16}
\end{align}
where
\begin{align}\label{q17}
\omega_\pm^2=\frac{1}{2}\bigg[\omega_m^2+\omega_b^2/\mu^2
\pm\sqrt{(\omega_m^2-\omega_b^2/\mu^2)^2
	+16G^2\mu\omega_m\omega_b}\bigg]
\end{align}
are respectively the eigen frequencies of the UBP and LBP with annihilation operators $d_+$ and $d_-$. As the UBP is always stable because $\omega_+^2>0$, so here we do not discuss it. Below we focus on the behavior of the LBP with the system parameters. To show this, we plot the normalized $\omega_-/\omega_m$ as functions of the normailized coupling strength $G/\omega_m$ and $\mu$ with $\omega_b/\omega_m=4$ in Fig.~\ref{fig2}.  We find $\omega_-^2=0$ gives rise to a CP $G/\omega_m=G_c/\omega_m=\sqrt{\omega_m\omega_b}/2=1$ in Fig.~\ref{fig2}(a). When $G>G_c$, $\omega_-^2>0$, otherwise $\omega_-^2<0$. Moreover, With increasing $G>G_c$, the LBP behaves like the normal quasi-particle with the positive frequency. From the view of the parameter $\mu$, the opposite behavior for the LBP can be obtained in Fig.~\ref{fig2}(b). Specifically, $\omega_-^2>0~(\omega_-^2<0)$ when $\mu<1~(\mu>1)$. Due to this critical behavior for the LBP, single-photon optomechanical coupling strengths can be enhanced beyond the low-excitation approximation (i.e., $\xi_1=1$ or $G<G_c$).
\begin{figure}
	\center
	\includegraphics[scale=0.45]{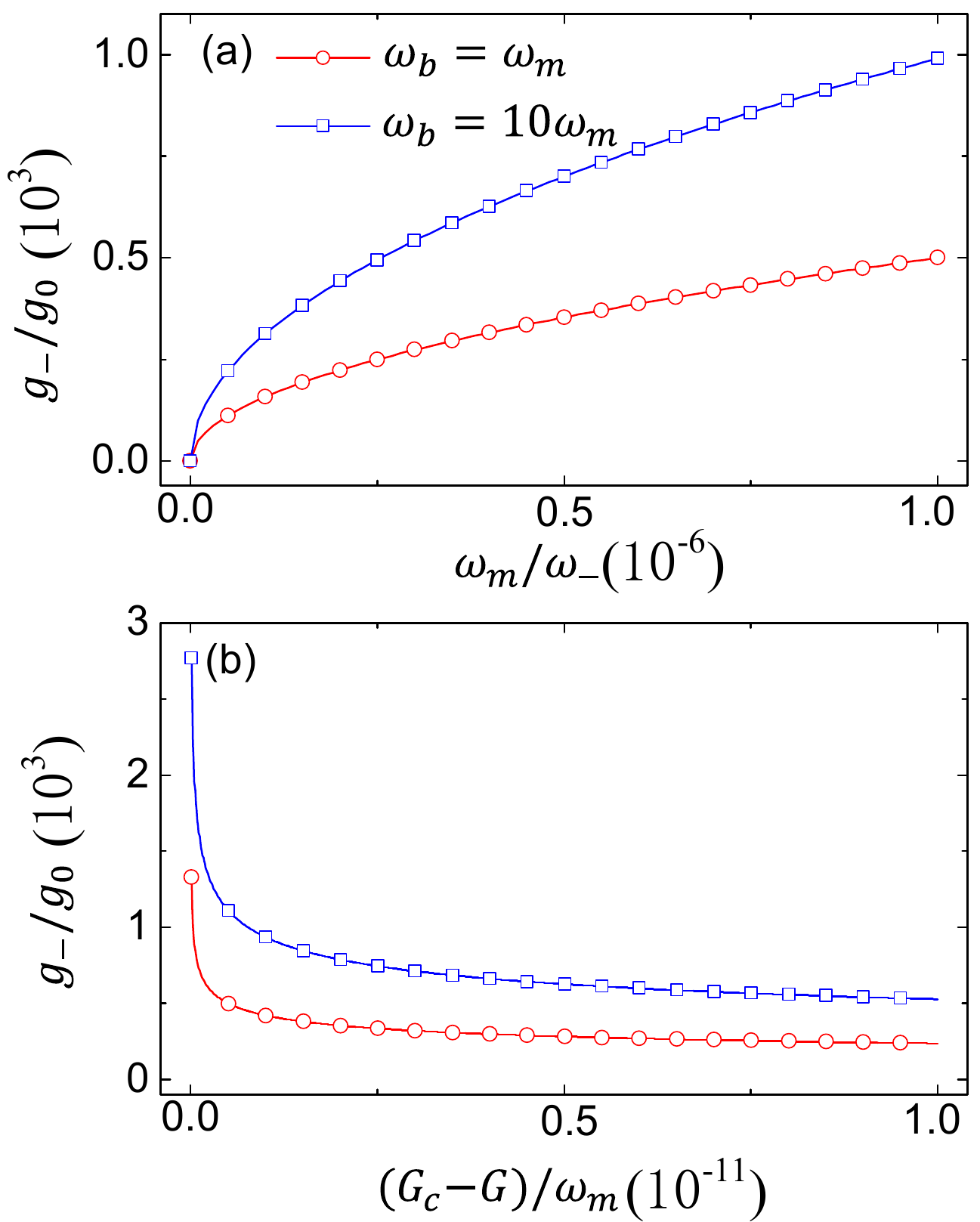}
	\caption{The normalized single-photon optomechanical oupling strengths of the cavity coupled to the LBP as functions of the normalized parameters $\omega_m /\omega_-$ and $(G_c-G)/\omega_m$ for $\omega_b/\omega_m=1$ and $10$.}\label{fig3}
\end{figure}

\section{Enhancement of optomechanical coupling at the single-photon level with quantum criticality}

In this part, we focus on enhancing the single-photon optomechanical coupling strength with the critical property of the system governed by Eq.~(\ref{q13}). To realize it, we couple the nanobeam to a(n) microwave (optical) cavity with a movable mirror, forming a cavity optomechanical system [see Fig.~\ref{fig1}(c)]. Thus, the total Hamiltonian of the hybrid system can be written as
\begin{align}
H_{\rm total}=H_{\rm OM}+H_{\rm NMP},\label{q18}
\end{align}
where $H_{\rm OM}=\omega_a a^\dag a-g_0 a^\dag a(b^\dag+b)$ is the Hamiltonian of the cavity optomechanical system. $\omega_a$ is the frequency of the cavity with annihilation (creation) operator $a$ $(a^\dag)$, $g_0$ is the single-photon optomechanical coupling strength, and $H_{\rm NMP}$ is given by Eq.~(\ref{q5}). In the polariton presentation, the phonon operators $b$ and $b^\dag$ can be expressed in terms of the polariton operators $d_\pm$ and $d_\pm^\dag$. Thus, Eq.~(\ref{q18}) can be written as
\begin{align}
H_{\rm OP}=&\omega_a a^\dag a+\omega_+ d_+^\dag d_++\omega_- d_-^\dag d_-\nonumber\\
&-g_+a^\dag a(d_+^\dag+d_+)+g_-a^\dag a(d_-^\dag+d_-),\label{q19}
\end{align}
where
\begin{align}
{g}_{+}=&g_0\sqrt{\omega_m/\omega_+}\cos\theta,\nonumber\\
{g}_{-}=&g_0\sqrt{\omega_m/\omega_-}\sin\theta,\label{q20}
\end{align}
with $\tan(2\theta)=2\omega_m\omega_b/(\omega_m^2-\omega_b^2/\mu^2)$,
are respectively the effective optomechanical coupling strengths bentween the cavity field and the UBP and LBP given by Eq.~(\ref{q13}) or Eq.~(\ref{q16}). Now we operate the subsystem of the NVE and the nanobeam around the CP, i.e., $G\rightarrow G_c$. This leads to $\omega_-\rightarrow 0$ and $\omega_+\rightarrow\sqrt{\omega_m^2+\omega_b^2}$, so $\omega_-\ll\omega_+$. Moreover, $\cos\theta\rightarrow\omega_m/\omega_+$ and $\sin\theta\rightarrow\omega_b/\omega_+$ when $\omega_-\rightarrow0$. In general, the frequency of the spin is larger than the frequency of the phonon in the nanobeam, that is, $\omega_b>\omega_m$. This causes $\sin\theta>\cos\theta$ in Eq.~(\ref{q20}). Based on above two points, we have $g_-\gg g_+$. To fully neglecting the coupling between the UBP and the cavity field, $\omega_b\gg\omega_m$ is expected. Therefore, near the CP, Eq.~(\ref{q19}) can be reduced to
\begin{align}
H=\omega_a a^\dag a+\omega_-d_-^\dag {d}_-
+g_- a^\dag a({d}_-^\dag+{d}_-).\label{q21}
\end{align}
To estimate $g_-$, we choose $\omega_-=2\pi\times10$ Hz, $\omega_m=10^6\omega_-\sim10$ MHz and $\omega_b\sim 10\omega_m$, which leads to $g_-\sim10^3 g_0$. This shows that the effective optomechanical coupling $g_-$ between the cavity field and the LBP is approximately enhanced by {\it three orders of magnitude} of the original optomechanical coupling strenght $g_0$ in our proposed hybrid system. This shows the enhanced single-photon optomecahnical coupling strength $g_-$ can be comparable with or exceed the decay rate of the cavity for most optomechanical experiments, i.e., $g_-\geq \kappa$. Thus, the enhanced single-photon optomechanical coupling can enter the {\it strong coupling regime}. To convinced this, the effective cooperativity $C_{\rm eff}=g_-^2/\kappa\gamma$~\cite{Li-2020,Qin-2018} is introduced for describing the strong coupling, and we find $C_{\rm eff}/C_0=10^6$, which indicates that the effective cooperativity is six orders of magnitude of the original cooperativity $C_0=g_0^2/\kappa\gamma$ at the vicinity of the CP. With above parameters, the critical coupling is about $G_c=\sqrt{\omega_b\omega_b}/2\sim1.6\omega_m$. To predict this CP, the total number of spins can be estimated as $N\sim10^{12}$ by using the accessible $g\sim 16$ Hz~\cite{Zhou-2017,Kubo-2010}. Actually, the value of $g$ can be further increased to kilohertz range by reducing dimensions of the mechanical oscillator~\cite{Babinec-2010}, which indicates that the required $N$ can be greatly reduced. At present, strong spin-phonon coupling via strain force has been experimentally achieved\cite{Ovartchaiyapong-2014,Teissier-2014}. Also, the NV spin can be
initialized to the $|m=0\rangle$ state via optical pumping with a $532$ nm laser, then microwave fields can be used to tune the population of the NV spin in the state $|m=-1\rangle$ or $|m=+1\rangle$~\cite{Ovartchaiyapong-2014}. Based on these, the critical property of the ensemble-phonon subsystem can be experimentally realized, which indicates that our proposal is feasible within the current technology.

In Fig.~\ref{fig3}, we also plot the normalized single-photon optomechanical coupling strength as functions of the normalized $\omega_m/\omega_-$ and $(G_c-G)/\omega_m$ for different $\omega_b/\omega_m$. From Fig.~\ref{fig3}(a), we see that the effective single-photon optomechanical coupling strength $g_-$  increases monotonously with increasing $\omega_m/\omega_-$, that is, the small $\omega_-$ leads to large $g_-$ for fixed $\omega_m$. However, $g_-$ decreases monotonously with increasing parameter $(G_c-G)/\omega_m$ describing the difference between $G_c$ and $G$ in Fig.~\ref{fig3}(b). When $G$ approaches $G_c$, $g_-$ can be greately enhanced. This enhancement can be up to three orders of magnitude of $g_0$, and we can obtain the larger $g_-$ by increasing the ratio of $\omega_b$ and $\omega_m$ in principle [see both Figs.~\ref{fig3}(a) and \ref{fig3}(b)].

The enhanced single-photon optomechanical coupling in Eq.~(\ref{q21}) can be used to produce strong Kerr nonlinearity, via the unitary transformation $U=\exp[-(g_-/\omega_-) a^\dag a({d}_-^\dag-{d}_-)]$, as
\begin{align}
H_{\rm K}=\omega_a a^\dag a+\chi (a^\dag a)^2,
\end{align}
where $\chi=g_-^2/\omega_-=g_0\sqrt{\omega_m/\omega_-^3}\sin\theta$. Obviously, the Kerr coefficient can be very strong due to $\omega_-\rightarrow 0$ at the CP. This Kerr has been investigated to generating Schr\"{o}dinger cat and studying photon blockade effect~\cite{Lu-2013}.

\section{Conclusion}
In summary, we have proposed a proposal to realize a strong single-photon optomechanical coupling in a hybrid spin-optomechanical system.  Different from previous low-excitation schemes, here we consider the spins are highly excited when the ensemble-phonon coupling exceeds a critical value. This coupling hybridizes the NVE and the phonon forming two polaritons. The LBP can have critical property when the NVE-phonon coupling approaches the critical value. Owning to this critical property, the coupling between the optomechanical cavity and the UBP can be fully suppressed, while the coupling between the cavity and the LBP is greatly improved. With  accessible parameters, the enhanced coupling can be about three orders of magnitude of the original value, so it can be in the strong coupling regime. This strong optomechanical coupling can be used to achieving strong Kerr nonlinearity, which can be employed to produce multi-component cat state~\cite{Lu-2013}. Our proposal provides a potential way to study quantum nonlinear optics in weakly coupled optomechanical systems. 

\section*{ACKNOWLEDGMENTS}
	This paper is supported by the National Natural Science Foundation of China (Grants No.~11804074, No.~11904201 and No.~12104214), and the Natural Science Foundation of Hunan Province of China (Grant No. 2020JJ5466).


\end{document}